%% file: 0Main.tex
\documentclass[fleqn,usenatbib]{mnras}

\usepackage{newtxtext,newtxmath}

\usepackage[T1]{fontenc}

\DeclareRobustCommand{\VAN}[3]{#2}
\let\VANthebibliography\thebibliography
\def\thebibliography{\DeclareRobustCommand{\VAN}[3]{##3}\VANthebibliography}

\usepackage{graphicx}	
\usepackage{amsmath}	
\usepackage{tabularx}
\usepackage{appendix}
\usepackage{enumitem}
\usepackage[table,x11names]{xcolor}
\usepackage{multirow}

\newcommand{\comment}[1]{}
\newcommand{\M}{M_{\odot}}
\newcommand{\tsfh}{\tau_{\mathrm{SFH}}}
\newcommand{\tstar}{\tau_{*}}
\newcommand{\yocc}{y\mathrm{_{\rm O}^{CC}}}

\newcommand{\yhecc}{y\mathrm{_{^3He}^{CC}}}
\newcommand{\yheagb}{y\mathrm{_{^3He}^{AGB}}}

\newcommand{\rgal}{$R_{\mathrm{gal}}$}
\newcommand{\zo}{Z_{\mathrm{O}}}
\newcommand{\zosun}{Z_{\mathrm{O}, \odot}}

\newcommand{\dely}{\Delta Y}

\newcommand{\hef}{^4\rm{He}}
\newcommand{\het}{^3\rm{He}}
\newcommand{\vice}{{\tt VICE}}
\newcommand{\ytsun}{Y_{3, \odot}}
\newcommand{\kms}{{\,{\rm km}\,{\rm s}^{-1}}}
\newcommand{\gyr}{\rm{Gyr}}

\title[GCE of $^3\rm{He}$]{Constraints on the Galactic Chemical Evolution of $^3$He}

\author[M. K. Weller et al.]{
Miqaela K. Weller,$^{1}$\thanks{E-mail: weller.133@osu.edu}
David H. Weinberg$^{1, 2}$
\\
$^{1}$Department of Astronomy, The Ohio State University, 140 West 18th Avenue, Columbus, OH 43210, USA\\
$^{2}$Center for Cosmology and AstroParticle Physics, The Ohio State University Columbus, OH 43210, USA
}

\date{Accepted XXX. Received YYY; in original form ZZZ}

\pubyear{2026}

\begin{document}
\label{firstpage}
\pagerange{\pageref{firstpage}--\pageref{lastpage}}
\maketitle

\begin{abstract}
We examine the galactic chemical evolution (GCE) of $^3\rm{He}$ in one-zone and multi-zone models, with particular attention to the stellar yields and GCE parameters that can reproduce both the protosolar $^3\rm{He}$ abundance and recent gas-phase $^3\rm{He}/^4\rm{He}$ measurements in the Orion nebula. Published stellar models indicate negligible net $^3\rm{He}$ production by massive stars, while the predicted yields from asymptotic giant branch (AGB) stars are metallicity-dependent and span a range of $\sim 2.5$ depending on the extra mixing processes incorporated in the stellar models. The dominant contribution to $^3\rm{He}$ production comes from $1-2\ M_\odot$ stars, making $^3\rm{He}$ evolution slow compared to other AGB elements and to Fe enrichment from Type Ia supernovae. We constrain our GCE models to reproduce the observed [O/H] in the interstellar medium, and our fiducial models adopt an empirically motivated IMF-averaged oxygen yield $y_{\rm O} \approx 1.2\ Z_{\rm O, \odot}$. Even with the lowest of the AGB $^3\rm{He}$ yields, based on stellar models with rotational and thermohaline mixing, our GCE models tend to overpredict the protosolar and Orion $^3\rm{He}$ abundances; they require a slow onset of star formation and low star formation efficiency to come close to the observed values. With a higher oxygen yield, calibration to observed [O/H] implies stronger outflows, making it easier to reproduce the observed $^3\rm{He}$. Alternatively, the true $^3\rm{He}$ yield could be lower than that predicted by existing stellar models, suggesting that mixing in red giants is not yet fully captured. Future $^3\rm{He}$ measurements that probe higher metallicity environments could help distinguish these possibilities.
\end{abstract}




\input{1Intro}
\input{2TheoreticalYieldCalculations}
\input{3OneZone}
\input{4MultiZone}
\input{5Conclusions}

\section*{Acknowledgements}
We thank Liam Dubay for sharing his implementation of the two-infall model within {\tt VICE} ahead of publication. We also thank James Johnson and Ryan Cooke for useful discussion and for providing the revised Orion $\het$ abundance from \citet{Cooke2026}.
This work was supported by NSF grants AST-1909841 and AST-2307621, by NASA grant 80NSSC24K0637, and by STScI award HST-GO-17303.002-A.

\section*{Data Availability}

{\tt VICE} is a publicly available code and the simulations can be
generated using the described parameters. Because \vice\ does not include ionic abundances, we use a different element to track $\het$, namely Au.



\bibliographystyle{mnras}
\bibliography{0Main}



\begin{appendices}
\input{6Appendix}
\end{appendices}

\bsp	
\label{lastpage}
\end{document}

%% file: 1Intro.tex
\section{Introduction}
\label{sec: intro}
$\het$ is the rarer, lighter isotope of helium. Like $\hef$, it was produced in the Big Bang, but in much smaller quantities, with a primordial number ratio $\het/\hef = 1.257 \times10^{-4}$ predicted by Big Bang Nucleosynthesis \citep[BBN; see, e.g.,][]{Pitrou21, Yeh2021}. Because of its low abundance, it is observationally challenging to disentangle $\het$ from $\hef$, but not impossible. The first $\het$ abundance measurements came from the combination of $\rm{D}/\hef$, $\rm{D/H}$, and $\het/\hef$ observations in the Sun \citep{Reeves1974}, H\,\textsc{ii} regions \citep{Predmore1971}, and young clusters \citep{Roger1971}. These were in apparent contradiction with each other, making the source of $\het$, i.e., stellar or cosmological, difficult to determine. Early theoretical models ascribed most $\het$ production to low mass stars that enriched the interstellar medium (ISM) during their final evolutionary stages \citep[e.g.,][]{Rood1976}. Interest in testing these models and using $\het$ as a probe of BBN motivated further studies exploring the chemical evolution of $\het$, with models often overproducing the protosolar abundance \citep[e.g.,][]{Flam1994, Galli1995, Dearborn1996}.

One possible solution to this overproduction is to adjust the yields of $\het$ to lower values, going beyond the predictions of standard stellar evolution models that mix stellar layers only through convection. Invoking ``extra mixing'' processes \citep[e.g.,][]{Charbonnel1998}, particularly rotational and thermohaline mixing \citep[e.g.,][]{Eggleton2006, Eggleton2008}, reduces the production of $\het$ because more of the $\het$ in a star's envelope is cycled to hotter interior zones where it can be fused to $\hef$. Thermohaline mixing, commonly known as the salt-finger instability, reduces the $\het$ abundance in stars through a mean-molecular weight inversion caused by the $\het$ ($\het$, 2p) $\hef$ reaction in the hydrogen burning shell on the Red Giant Branch. Mixing arises in zones that are convectively stable because heavier material (high $\mu$) sits on top of lighter material (low $\mu$).

Some measurements of planetary nebulae (PNe) \citep[e.g.,][]{Balser1999, Balser2006, Ramirez2016}, on the other hand, showed abundances consistent with standard stellar evolution models, suggesting that not all stars undergo extra mixing. To account for this possibility, \citet{Lagarde2011} constructed models in which a small fraction (4\%) of stars follow standard stellar evolution predictions and the rest undergo extra mixing. However, \citet{Bania2021} and \citet{Balser2022} have since suggested that these PNe detections should be treated as upper limits, removing the evidence that some stars do not undergo extra mixing.

In this paper, we construct galactic chemical evolution (GCE) models for $\het$ incorporating the \citeauthor{Lagarde2011}\ (\citeyear{Lagarde2011}, hereafter L11) stellar yields, with methods similar to those in our recent study of $\hef$ evolution \citep{Weller2024}. We find that $\het$ exhibits distinctive behavior because its production is dominated by long-lived, low mass stars ($1-2 M_\odot$), making $\het$ enrichment slow compared to $\hef$ or even to $s$-process elements that have a large contribution from asymptotic giant branch (AGB) stars.  Our primary observational targets are the protosolar $\het$ mass fraction $Y_{3,\odot} = 3.38 \times 10^{-5}$ \citep{Mahaffy1998} and the interstellar medium $\het/\hef$ measurement in Orion by \cite{Cooke22}, as recently refined by \cite{Cooke2026}, implying $Y_{3, \rm Orion} = \left(3.78 \pm 0.14\right) \times 10^{-5}$.  We also consider ISM $\het/H$ measurements by \cite{Balser2018}, which have lower precision but span a wider range of Galactic environments. Relative to the models of \cite{Cooke22}, we investigate a wider variety of GCE parameters to understand how they change $Y_3$ predictions and what values are required to explain the protosolar and Orion abundances, given the L11 yields and the primordial $Y_{3,\rm{P}} \approx 2.45\times 10^{-5}$ predicted by BBN \citep[e.g.,][]{Pitrou21, Yeh2021}. The ISM D/H and $\het/\rm{H}$ ratios provide sensitive diagnostics for the importance of outflows or radial gas flows in GCE and, when combined with the ISM oxygen abundance, for the overall scale of nucleosynthetic yields \citep{WAF, Weinberg24, Johnson25}. The interplay between $Y_3$ predictions and the nucleoynthetic yield scale is also an important theme of our study.

This paper is organized as follows. In Section \ref{sec: yields}, we explore the production, or destruction, of $\het$ in both low mass and high mass stars to define a population averaged yield. In Section \ref{sec: onezone}, we run one-zone chemical evolution models to try and reproduce updated observations. In Section \ref{sec: multizone}, we run multi-zone models with the adjusted parameters from Section \ref{sec: onezone}. Finally, in Section \ref{sec: conclusions}, we discuss our main conclusions. To be explicit, in the general discussion of elements, we use $\het$ and $\hef$, but in terms of abundance values, we use different nomenclature that is defined when first used.

\begin{figure*}
    \centering
    \centerline{\includegraphics[width=0.85\paperwidth]{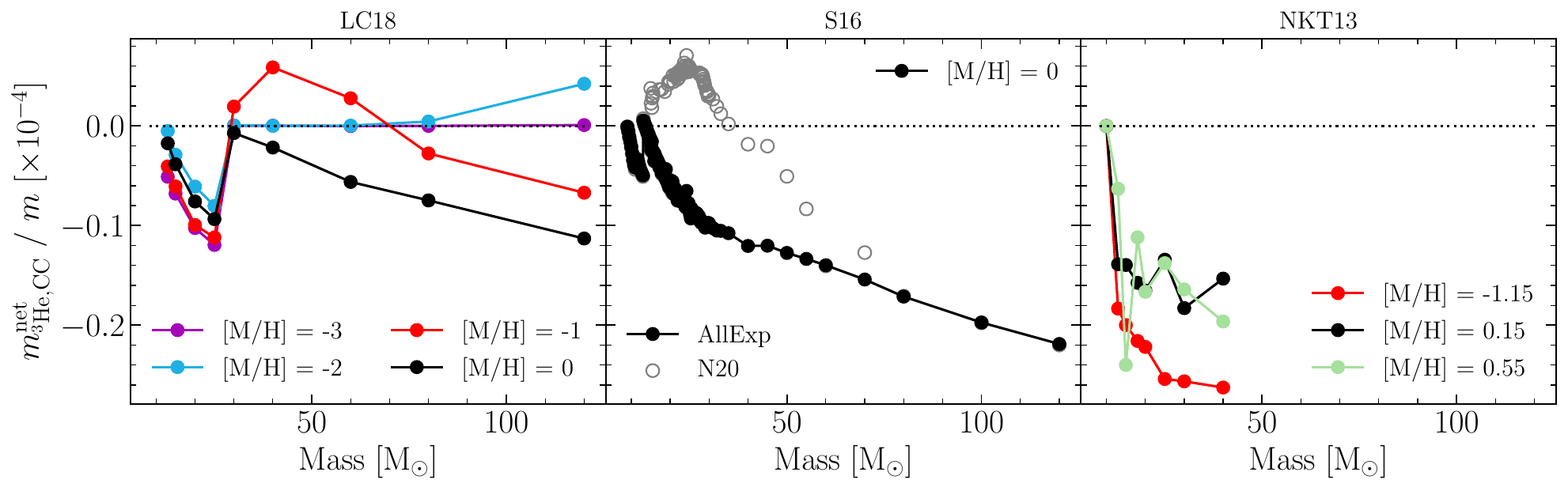}}
    \caption{Fractional net $\het$ yields of massive stars as a function of stellar mass, based on the stellar evolution and supernova models of \citet{LC18}, \citet{S16}, and \citet{NKT13}. Consistent color-coding is used between panels to denote similar metallicities reported by the different studies. Black points (AllExp) in the S16 panel show calculations with forced explosions of all progenitors as described by \citet{Griffith21}. Gray open circles (note that this is for [M/H] = 0 as well) show the original S16 results for their N20 central engine, where many progenitors collapse to black holes. Points below the dotted lines represent a net sink of $\het$, meaning that the star consumed more $\het$ than it produced.}
    \label{fig: ccnet}
\end{figure*}

%% file: 2TheoreticalYieldCalculations.tex
\section{Theoretical Yield Calculations}
\label{sec: yields}

In the approximation of instantaneous enrichment and recycling, the evolution of the mass of $\het$ in the ISM can be written as
\begin{equation}
    \label{enrichment}
    \dot{M}_{\mathrm{^3He}} = \yhecc\dot{M}_{*} + \yheagb\dot{M}_{*} + rY_3\dot{M_{*}} + Y_{3, \rm P}\dot{M}_{\mathrm{acc}} - Y_3\dot{M_{*}} - \eta Y_3\dot{M_{*}},
\end{equation}
where $Y_3 \equiv M_{\mathrm{^3He}}/M_{\rm gas}$ is the mass fraction in the ISM. While the instantaneous approximation is not very accurate for $\het$, as we show below, Equation \ref{enrichment} is conceptually useful for understanding the sources and sinks of $\het$. The quantities $\yhecc$ and $\yheagb$ represent the \textit{net} yield of $\het$ from massive stars and asymptotic giant branch (AGB) stars, respectively, per unit mass of star formation. We denote the massive star yield CC because of the connection to core-collapse supernovae (CCSN), though for $\het$ much of the massive star yield is released in stellar winds. We compute these net yields assuming a \citet{Kroupa} initial mass function (IMF) and the stellar yields described in Sections \ref{ccsn_yields} and \ref{agb_yields}, with the boundary between massive star and AGB enrichment at $m = 8 \M$. The third term represents $\het$ that is returned by the envelopes of stars at their birth abundance, without new synthesis or destruction. For a \citet{Kroupa} IMF, $r \approx 0.4$, with roughly equal contributions above and below $8 \M$. Accretion of gas, assumed to have the primordial abundance $Y_{3, \rm P} \approx 10^{-5}$, is another source of ISM $\het$. The two sink terms in Equation \ref{enrichment} represent the depletion of ISM $\het$ into new stars and the ejection of ISM in a galactic wind with mass-loading factor $\eta = \dot{M}_{\mathrm{out}}/\dot{M}_{*}$.

For consistency with Equation \ref{enrichment}, we define the net yield of a star of birth mass $m$ by
\begin{equation}
    \label{yield}
    m\mathrm{_{^3He}^{net}} = m\mathrm{_{^3He}^{gross}} - Y_{3, i} (m - m_{\mathrm{rem}}),
\end{equation}
where $Y_{3, i}$ is the initial (birth) mass fraction, $m_{\rm rem}$ is the mass of the stellar remnant, and $m\mathrm{_{^3He}^{gross}}$ is the total $\het$ mass returned to the ISM at stellar death. The net yield can be negative if $m\mathrm{_{^3He}^{gross}}$ is lower than $Y_{3, i} (m - m_{\mathrm{rem}})$, implying that the birth and death of the star results in a net loss of ISM $\het$. 

\subsection{CCSN and Massive Star Winds}
\label{ccsn_yields}
Due to the temperatures of high mass stars, the CNO bi-cycle dominates the interior hydrogen fusion process. Even though $\het$ does not play a central role in this cycle, any $\het$ will be efficiently destroyed in the hotter regions of a star. In Figure \ref{fig: ccnet}, we plot the net yield of $\het$ as a function of zero-age main-sequence mass for 3 different massive star evolution studies, \citet[][LC18]{LC18}, \citet[][S16]{S16}, and \citet[][NKT13]{NKT13}. We color-code by similar metallicities across panels. For \citet{S16}, we plot the yield for two different central engines, which affect the explosion landscape of massive stars. Overall, we find that the majority of massive stars do not produce any new $\het$ and instead are a net sink. However, stars with masses of $30 \M \lesssim m \lesssim 70\M$ at [M/H] = -1 from LC18 and $20 \M \lesssim m \lesssim 40\M$ at [M/H] = 0 from S16 for the N20 central engine are net producers. This is likely a balance of intermediate mass loss and opacity effects making the stellar interior cooler so that $\het$ can survive.

For computing the IMF-averaged yield, we define
\begin{equation}
    \label{frac}
    \yhecc = \frac{\int_{m\mathrm{_{SN}}}^{m\mathrm{_{max}}} [E(m) m\mathrm{_{^3He, exp}^{gross}} + m\mathrm{_{^3He, wind}^{gross}} - Y_{3, i}(m - m_{\mathrm{rem}})] \frac{dN}{dm} dm}{\int_{m\mathrm{_{min}}}^{m\mathrm{_{max}}} m \frac{dN}{dm} dm}.
\end{equation}
The numerator is similar to Equation \ref{yield}, but for massive stars we separate out an explosive and wind component for gross yields. We define $E(m)$ as the explodability function, where $E(m) = 1$ defines an explosion that releases material and $E(m) = 0$ signifies black hole formation in which no material escapes. We adopt $m_{\mathrm{min}}$ = 0.08 $\M$ and $m_{\mathrm{max}}$ = 120 $\M$ as the limits of the IMF and $m_{\mathrm{SN}}$ = 8 $\M$ as the minimum mass for a supernova progenitor. A similar equation describes $\yocc$, the IMF-averaged yield for oxygen, replacing $Y_{3,\ i}$ with the oxygen mass fraction $Z_{\mathrm{O},\ i}$.

\subsection{AGB Stars}
\label{agb_yields}
Instead of the CNO bi-cycle, low mass stars undergo the pp-chain, in which $\het$ plays a critical role in producing $\hef$. Because of this, there can be a slight buildup of $\het$ that can then escape the star during shell flashes in the AGB phase. However, extra mixing processes can result in $\het$ burning when envelope material is brought to the more central regions of a star. \citet{Lagarde2011} explore such processes in their stellar models.

\begin{figure*}
    \centering
    \centerline{\includegraphics[width=0.85\paperwidth]{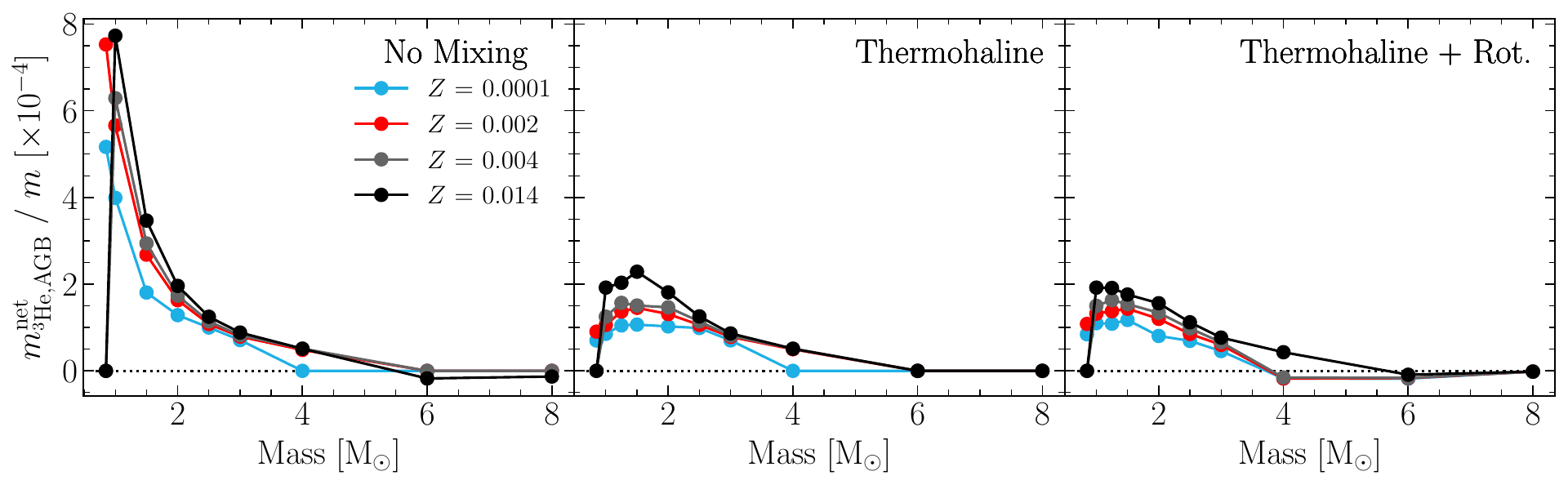}}
    \caption{Fractional net $\het$ yields from \citet{Lagarde2011} for models with no mixing and models that include extra mixing processes (i.e., thermohaline and rotational mixing) as a function of mass and metallicity. If the yield at 8 $\M$ is not reported, we adopt the yield at 6 $\M$ in lieu of linearly extrapolating. Like with Figure \ref{fig: ccnet}, the black dotted line denotes zero $\het$ production and we have again displayed metallicities with similar colors for ease of comparison.}
    \label{fig: agbnet}
\end{figure*}

In Figure \ref{fig: agbnet}, we show the L11 yields using standard stellar evolution (i.e., no mixing) and models with extra mixing processes at different metallicities. From these yields, it is evident that AGB stars are a net producer of $\het$, especially in the models with no mixing. However, if processes like thermohaline and rotational mixing are present, this leads to more $\het$ being destroyed inside of stars. Figure \ref{fig: agbnet} also implies that most $\het$ production occurs in low mass stars, with birth mass $m \approx 1-3 \M$.\footnote{For contribution to the IMF averaged yield, one should multiply these curves by $m$ to convert fractional to total yields and by $m^{-2.3}$ to account for the IMF, a combined factor of $m^{-1.3}$.}

\begin{figure*}
    \centering
\centerline{\includegraphics[width=0.85\paperwidth]{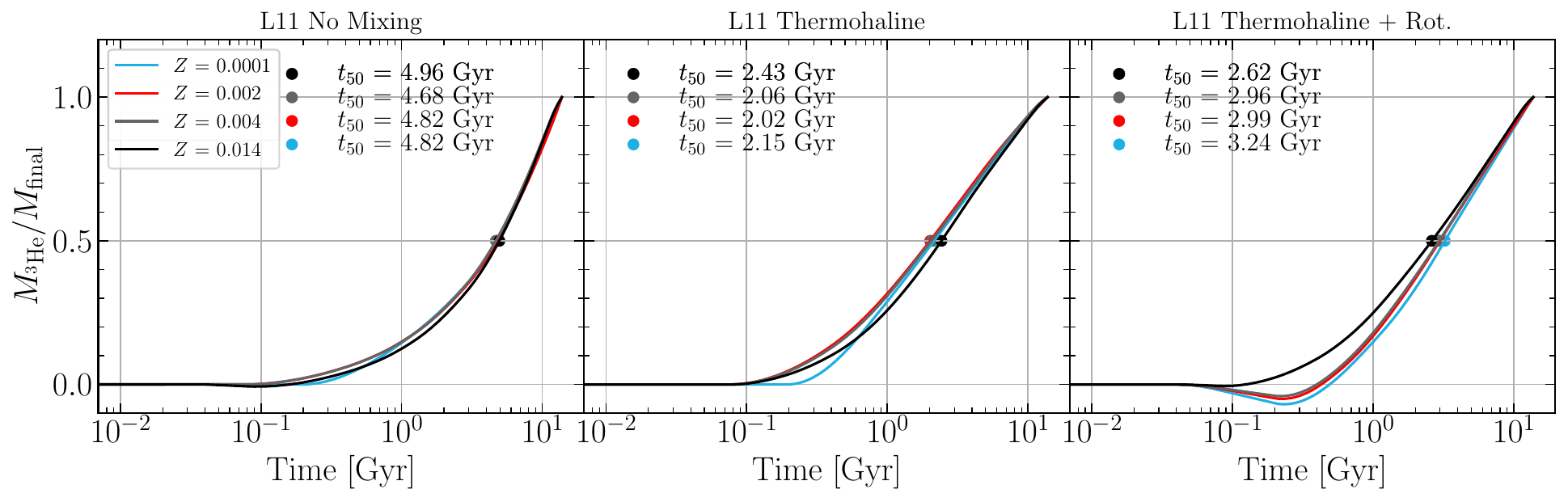}}
    \caption{Production of $\het$ from AGB stars only as a function of time for a population of stars born at $t = 0$. The markers indicate the time at which 50\% of the total $\het$ has been produced. This value is larger for the standard yield cases as 1 -- 2 $\M$ stars dominate the production.}
    \label{fig: agbdelay}
\end{figure*}

This behavior contrasts with that of $\hef$ and most neutron capture elements, for which the dominant producers have $m > 2\M$ (see, e.g., \citet{Johnson20} for Sr and \citet{Weller2024} for $\hef$). Figure \ref{fig: agbdelay} plots the net AGB $\het$ production vs. time for a single stellar population born at $t = 0$. We refer to this normalized evolution curve as the delay time distribution (DTD) for AGB $\het$ enrichment, analogous to the DTD for iron (Fe) production by Type Ia supernovae (SNIa). Because of the large contribution from low mass stars, the predicted time delays are quite long, with half of the AGB $\het$ production occurring in a time $t_{50} \sim 4.6\ \rm{Gyr}$ for the no mixing yields and $\sim 3\ \rm{Gyr}$ for the thermohaline + rotational mixing yields. These are much longer than typical estimates of median time delays for SNIa Fe production \citep[$\sim 1\ \rm{Gyr}$;][]{Dubay25} or AGB production of Sr \citep[$\sim 0.5\ \rm{Gyr}$;][]{Johnson20} or N \citep[$\sim 0.25\ \rm{Gyr}$;][]{Johnson23}.

For analytic models discussed below, we convert $t_{50}$ values to equivalent exponential timescales from the condition
\begin{equation}
    \label{half_production}
    \frac{\int_{t_{D}}^{t_{50}} e^{-t/\tau} dt}{\int_{t_{D}}^{\infty} e^{-t/\tau} dt} = 0.5,
\end{equation}
where $t_D \sim 50\ \rm{Myr}$ is the lifetime of an 8 $\M$ star \citep{Larson74}. Equation \ref{half_production} implies $\tau = (t_{50} - t_D) / \rm{ln}(2)$, which in combination with the results of Figure \ref{fig: agbdelay} gives $\tau_{\rm{no\ mixing}} = 7.1\ \rm{Gyr}$ and $\tau_{\rm{thm+rot}} = 3.71\ \rm{Gyr}$ at solar metallicity. Our numerical calculations below account for the time dependence of AGB enrichment and recycling.

\subsection{Summary}
\begin{figure}
    \centering
\centerline{\includegraphics[width=0.85\columnwidth]{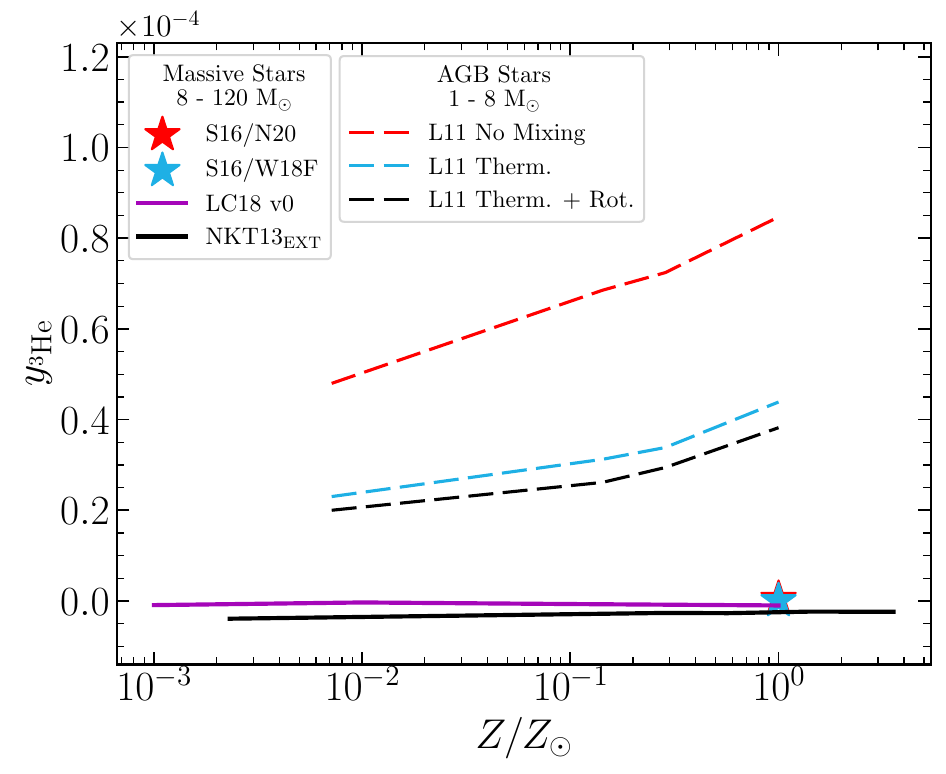}}
    \caption{Net IMF-averaged yields of massive stars (solid and markers) and AGB stars (dashed). The yields from massive stars are slightly negative and negligible compared to yields from AGB stars. Our preferred yield set, L11 Therm. + Rot., increases with metallicity by a factor of $\sim2$ over our metallicity range.}
    \label{fig: imf_yields}
\end{figure}

Figure \ref{fig: imf_yields} shows the dimensionless, IMF averaged $\het$ yields for massive and low mass stars, as a function of metallicity. The predicted $\het$ yields of massive stars are negligible relative to those of AGB stars, though we do include them in our models below using the tables of LC18, which include explosive and stellar winds for stars up to $120\ \M$. For a given mixing model, the L11 yields grow by a factor of $\sim2$ between $Z = 0.01\ Z_{\odot}$ and $Z_{\odot}$. At a given $Z$, the predicted yields of the thermohaline + rotational mixing models are a factor of $\sim2.5$ lower than those of standard (no extra mixing) models. We adopt the thermohaline + rotational mixing yields as our default below, since previous studies have reported overproducing $\het$, and we find that even these lower yields are difficult to reconcile with observed $\het$ abundances.

%% file: 3OneZone.tex
\section{One-zone Galactic Chemical Evolution}
\label{sec: onezone}

To craft a picture of the production and evolution of $\het$, we begin with one-zone chemical evolution models. These models, which can be computed analytically with some restrictions and numerically for more general cases, assume a fully mixed, chemically homogeneous gas reservoir that provides the fuel for star-formation. With the lives and deaths of stars, this reservoir is injected with the products of stellar nucleosynthesis.

All of our models include accretion of gas, which is assumed to have primordial composition, and they allow ejection of gas by galactic winds, which are assumed to have the same composition as the ISM. For numerical calculations, we use the Versatile Integrator for Chemical Evolution \citep[{\tt VICE};][]{Johnson20}\footnote{https://vice-astro.readthedocs.io/en/latest/index.html}. With the restriction of metallicity-independent yields, an exponential DTD, and instantaneous recycling (but not instantaneous enrichment), we can also compute analytic solutions with the formalism of \citet{WAF}, adapting their results for SNIa Fe production to AGB $\het$ production. Comparison of numerical and analytic results helps to show how different aspects of $\het$ production affect our results.

The key parameters of our one-zone models, in addition to yields, are the star formation efficiency (SFE) timescale $\tstar = M_{\rm gas} / \dot{M}_*$, the outflow efficiency $\eta = \dot{M}_{\rm out} / \dot{M}_*$, and the e-folding timescale $\tsfh$ of the star formation history (SFH), assumed to be an exponential with $\dot{M}_*(t) \propto e^{-t/\tsfh}$. Gas accretion occurs at the rate required to maintain the specified SFH. \citet{WAF} examine the effects of $\tstar$, $\eta$, and $\tsfh$ on O and Fe evolution. The characteristic timescale for O evolution (or for other elements dominated by massive star production) is the depletion time $\tau_{\rm{dep}} = \tstar/(1 + \eta - r)$, though evolution is more extended if $\tsfh$ begins to approach $\tau_{\rm dep}$. \citet{Weller2024} describe the evolution of $\hef$ in similar models.

We adopt $Y_{3, \odot} = 3.38\times10^{-5}$ for the protosolar $\het$ mass fraction based on the ratio $\het/\hef = 1.66\times10^{-4}$ measured by \citet{Mahaffy1998} in the atmosphere of Jupiter (and $Y_{4, \odot} = 0.2703$ from \citealt{Asplund2009}); the higher ratio in the solar wind \citep{Heber12} is affected by fusion of D into $\het$ during the Sun's pre-main sequence phase. For the protosolar oxygen abundance, we adopt $\zosun = 7.33\times10^{-3}$ based on the \citet{Magg22} photospheric abundance with a +0.04 dex correction for diffusion. Our fiducial oxygen yield is $\yocc = 0.009 = 1.23\ \zosun$, as in the $\hef$ evolution models of \citet{Weller2024}. This yield is 0.1 dex higher than the value advocated by \citet{Weinberg24} based on calibration to the mean CCSN Fe yield \citep{Rodriguez23}, but \citet{Weller2024} find that this increase is needed to reproduce the solar $\hef/\rm{O}$ ratio even with the lowest theoretical $\hef$ yields that they consider. Unless otherwise specified, we use the thermohaline + rotational mixing yield of L11, setting $Z/Z_{\odot} = \zo/\zosun$ for metallicity dependence.

\begin{figure}
\centering
    \includegraphics[width=0.9\columnwidth]{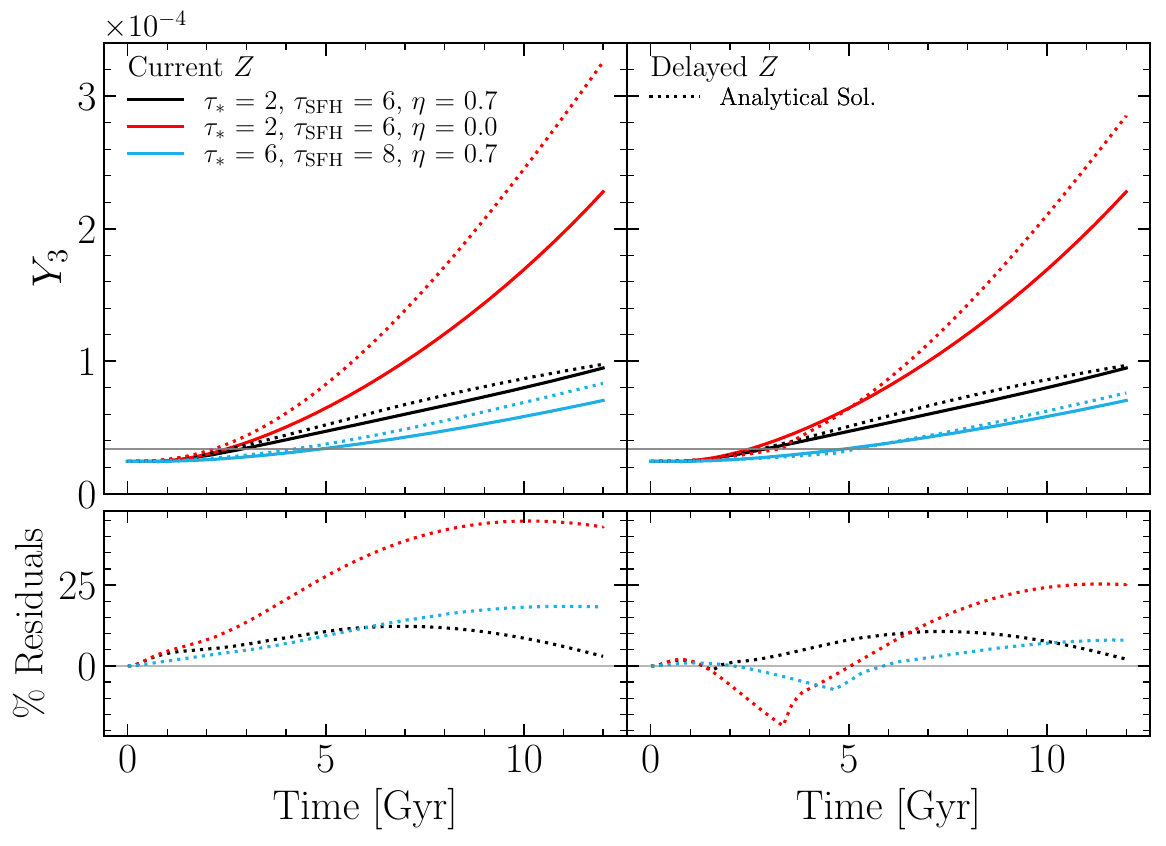}
    \caption{$Y_3$ as a function of time for one-zone models with exponentially declining SFH, for three different combinations of SFE timescale $\tstar$, SFH decline timescale $\tsfh$, and outflow efficiency $\eta$. Dotted curves in the upper panels show analytic solutions with the $\het$ yield computed for the metallicity at time $t$ (left) or time $t - t_{\rm dep}$ (right). Lower panels show residuals between analytic and numerical solutions.}
    \label{fig: cz_vs_dlz}
\end{figure}

\begin{figure*}
    \centering
\centerline{\includegraphics[width=0.85\paperwidth]{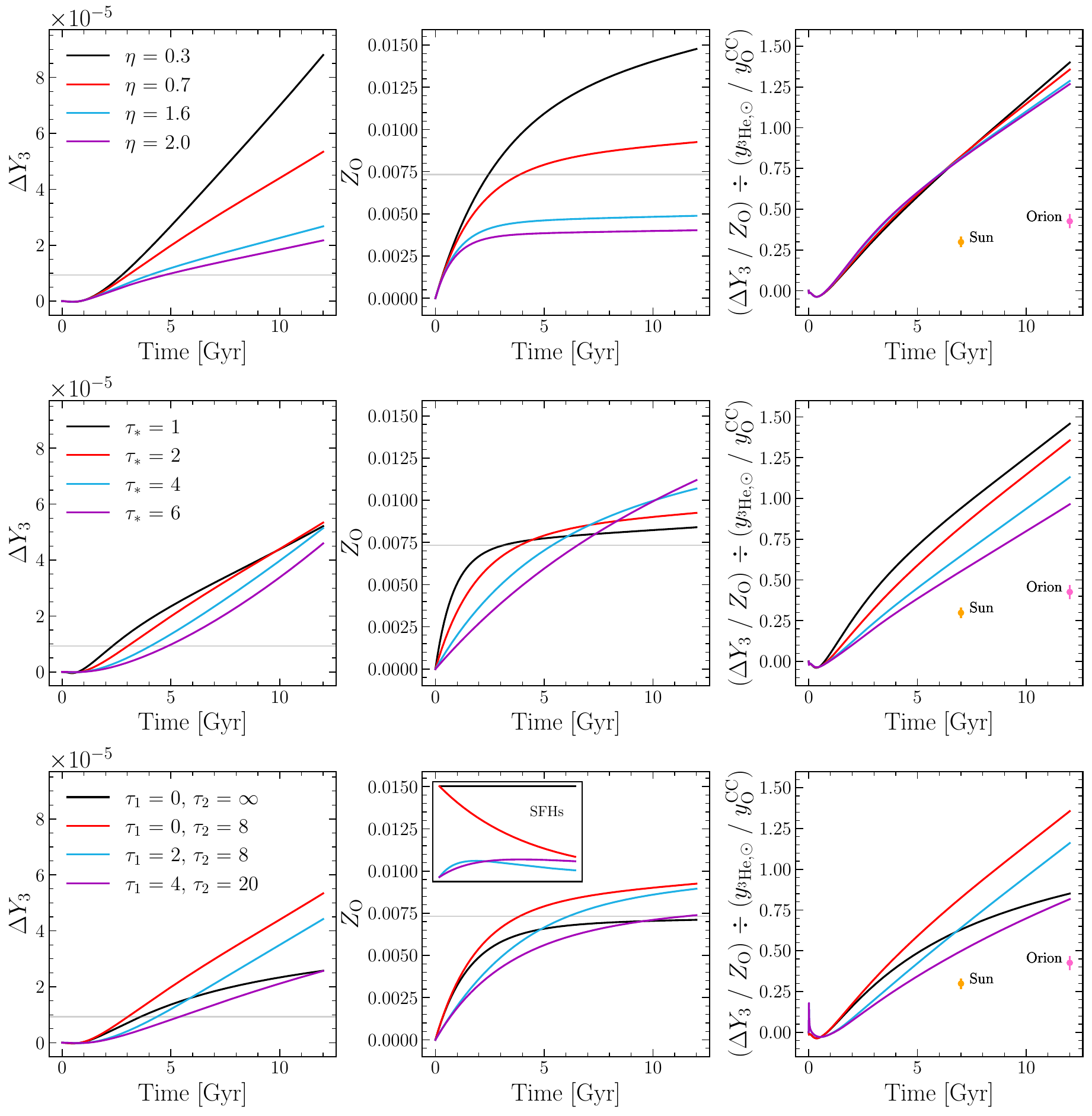}}
    \caption{Suite of numerical one-zone models plotting $\Delta Y_3$, $\zo$, and the ratio $\dely_3/\zo$ normalized by the yield ratio at solar metallicity with varying parameters, all assuming $y_{\rm O}^{\rm CC} = 0.009$. \textit{Top:} Changes in the outflow affect $\het$ and oxygen in similar ways. \textit{Middle:} Increasing the star-formation efficiency timescale affects $\het$ differently than oxygen because of the longer delay time. \textit{Bottom:} Changes to star formation history parameters have qualitatively similar effects to changes in $\tau_*$. Star formation histories in the top and middle rows have $\tau_1 = 0$, $\tau_2 = 8$, and SFHs for the bottom row are shown schematically in the middle panel.}
    \label{fig: oz_vary}
\end{figure*}

\subsection{The Impact of Outflows and Star Formation Efficiency}
\label{sec: outflows_sfe}

Figure \ref{fig: cz_vs_dlz} shows the evolution of $Y_3$ in three representative models with different combinations of $\eta$, $\tstar$, and $\tsfh$. With $\tstar = 2\ \rm{Gyr}$ \citep[a typical value for molecular gas;][]{Leroy2008, Sun22} and a moderately declining SFH, a value of $\eta = 0.7$ leads to roughly solar $\zo$ at late times. This model predicts $Y_3 \approx 10^{-4} \approx 3\ \ytsun$ at the present day (black solid curve). Setting $\eta = 0$ (no outflow, red solid curve) raises $Y_3$ substantially because newly produced $\het$ is not ejected from the ISM and because the ISM oxygen abundance is higher at each $t$, implying a higher AGB $\het$ yield. Increasing the SFE and SFH timescales, to $\tstar = 6\ \rm{Gyr}$ and $\tsfh = 8\ \rm{Gyr}$ (blue solid curve), leads to slower evolution and a lower final $Y_3$, but still substantially above the solar value. In all of our one-zone models, we adopt $t = 12\ \rm{Gyr}$ as the age of the disk, effectively assuming that disk formation and chemical evolution does not begin until $\approx 2\ \rm{Gyr}$ after the Big Bang. 

\citet{WAF} present analytic solutions for the evolution of an element with a metallicity-independent yield and an exponential enrichment DTD, which they apply to Fe enrichment from SNIa. Dotted curves in the left panels of Figure \ref{fig: cz_vs_dlz} show the application of this same solution (their Equation 53) to $\het$ evolution with the substitution $\tau_{\rm Ia} \rightarrow \tau_{\rm AGB} = 3.71\ \rm{Gyr}$ and the yield at time $t$ based on the oxygen abundance at time $t$ (computed from \citet{WAF}'s Equation 50). Specifically, we use this solution for $\dely_3 = Y_3 - Y_{3, \rm P}$, then add $Y_{3, \rm P}$ to compute $Y_3$. We provide the analytic equations for $\het$ adapted from \citet{WAF} in Appendix \ref{appendix:extras}. The analytic solutions overpredict $\het$, in part because they assume yields at the current metallicity, whereas the ISM at time $t$ is enriched by stars formed at earlier, lower-metallicity epochs with lower $\het$ yields. In the right panels of Figure \ref{fig: cz_vs_dlz}, we change the analytic solutions to use the metallicity at time $t - \tau_{\rm dep}$ when computing the yield (or the lowest $Z$ yield for $t < \tau_{\rm dep}$). While these analytic solutions are not accurate enough to replace the numerical calculations, they capture the models' qualitative behavior and parameter dependence, suggesting that many aspects of $\het$ evolution can be understood with the \citet{WAF} framework. \citet{Sanders25} shows that this analytic approach can be extended to the case of metallicity-dependent yields with the aid of Laplace transforms. We have not pursued this approach here, but we expect that the \citet{Sanders25} formalism could be combined with the yields of Figure \ref{fig: imf_yields} to give accurate analytic predictions of $\het$ evolution for smooth star formation histories. 

Figure \ref{fig: oz_vary} examines the evolution of $\dely_3$ (left), O (middle), and their ratio (right), for a wider array of models. In all panels, red curves show a fiducial model with $\tstar = 2\ \rm{Gyr}$, $\eta = 0.7$, and an exponential SFH with $\tsfh = 8\ \rm{Gyr}$. Raising (lowering) $\eta$ leads to lower (higher) $\dely_3$ and $\zo$, the expected impact of ISM losses in outflows (see Equation \ref{enrichment}). While oxygen approaches an equilibrium at late times, the $\dely_3$ curves rise steadily because of the long timescale of AGB enrichment. The ratio $\dely_3/\zo$ rises steadily with time in a way that is nearly independent of $\eta$. This steady rise of $\dely_3/\zo$ is quite different from the behavior of $\hef$ (see, e.g., Figure 12 of \citealt{Weller2024}), because the AGB enrichment of $\hef$ is dominated by higher mass stars with much shorter lifetimes.

We normalize the $\dely_3/\zo$ ratio by the corresponding ratio of yields $y_{\het, \odot}/\yocc$, since with other parameters held fixed, all $\zo(t)$ values are proportional to $\yocc$, and all $\dely_3(t)$ values are proportional to the overall normalization of $\het$ yields. We show the $\dely_3/\zo$ ratios for the Sun and the present-day measurement in Orion (\citeauthor{Cooke22}~\citeyear{Cooke22}, \citeyear{Cooke2026}), adopting $\zo = \zosun$ for the latter and normalizing with our fiducial yield choices of $\yocc = 0.009$ and $y_{\het, \odot} = 3.82 \times 10^{-5}$ (L11, thermohaline + rotational). All models in the top row overpredict the observed $\dely_3/\zo$ ratios by a substantial factor.

The second row shows the impact of varying $\tstar$. Lower SFE (longer $\tstar$) leads to slower evolution for both $\het$ and O, but the impact on $\het$ is larger because slow gas depletion combines with slow AGB enrichment, so the $\dely_3/\zo$ ratio is lower for longer $\tstar$. While a value $\tstar = 6\ \rm{Gyr}$ might be reasonable for the solar neighborhood at the present-day, it is much longer than the $\tstar = 2\ \rm{Gyr}$ found in regions where molecular gas dominates the ISM surface density \citep{Leroy2008, Sun22}, so it seems improbably high for earlier phases of the Milky Way's history when the disk was more gas rich. Even the $\tstar = 6\ \rm{Gyr}$ model overpredicts the observed $\dely_3/\zo$ values. 

\subsection{The Impact of Star Formation History and Oxygen Yield Normalization}
\label{sec: sfh_oxy}

The bottom row of Figure \ref{fig: oz_vary} examines variations in SFH, parameterized by $\dot{M_*}(t) \propto \left(1 - e^{-t/\tau_1}\right)e^{-t/\tau_2}$, all with $\eta = 0.7$ and $\tstar = 2\ \rm{Gyr}$. For $\tau_1 \ll \tau_2$, this SFH rises linearly at early times, reaches a maximum at $t \approx \tau_1$, and declines exponentially at late times with timescale $\tau_2$. The $(\tau_1, \tau_2) = (0, 8)$ model is our fiducial exponential SFH (red curves, identical to those in the upper rows). The $(\tau_1, \tau_2) = (0, \infty)$ model (black curves) corresponds to a constant SFR, which leads to lower $\zo$ and substantially lower $\dely_3$ at late times because maintaining constant SFR requires a higher accretion rate of (metal-free, primordial $Y_3$) gas at late times. The $(\tau_1, \tau_2) = (2, 8)$ model (blue curves) resembles the fiducial model at late times, but the slower SFR at early times leads to delayed enrichment in both $\zo$ and $\dely_3$. The $(\tau_1, \tau_2) = (4, 20)$ model (purple curves) combines a slow start with nearly constant SFR at late times, so it predicts the lowest $\dely_3$ and $\zo$ through most of its history, though it converges with the constant SFR model at late times. This model still overproduces the observed $\dely_3/\zo$ values, but it comes the closest of the models in this Figure.

\begin{figure*}
    \centering
\centerline{\includegraphics[width=0.7\paperwidth]{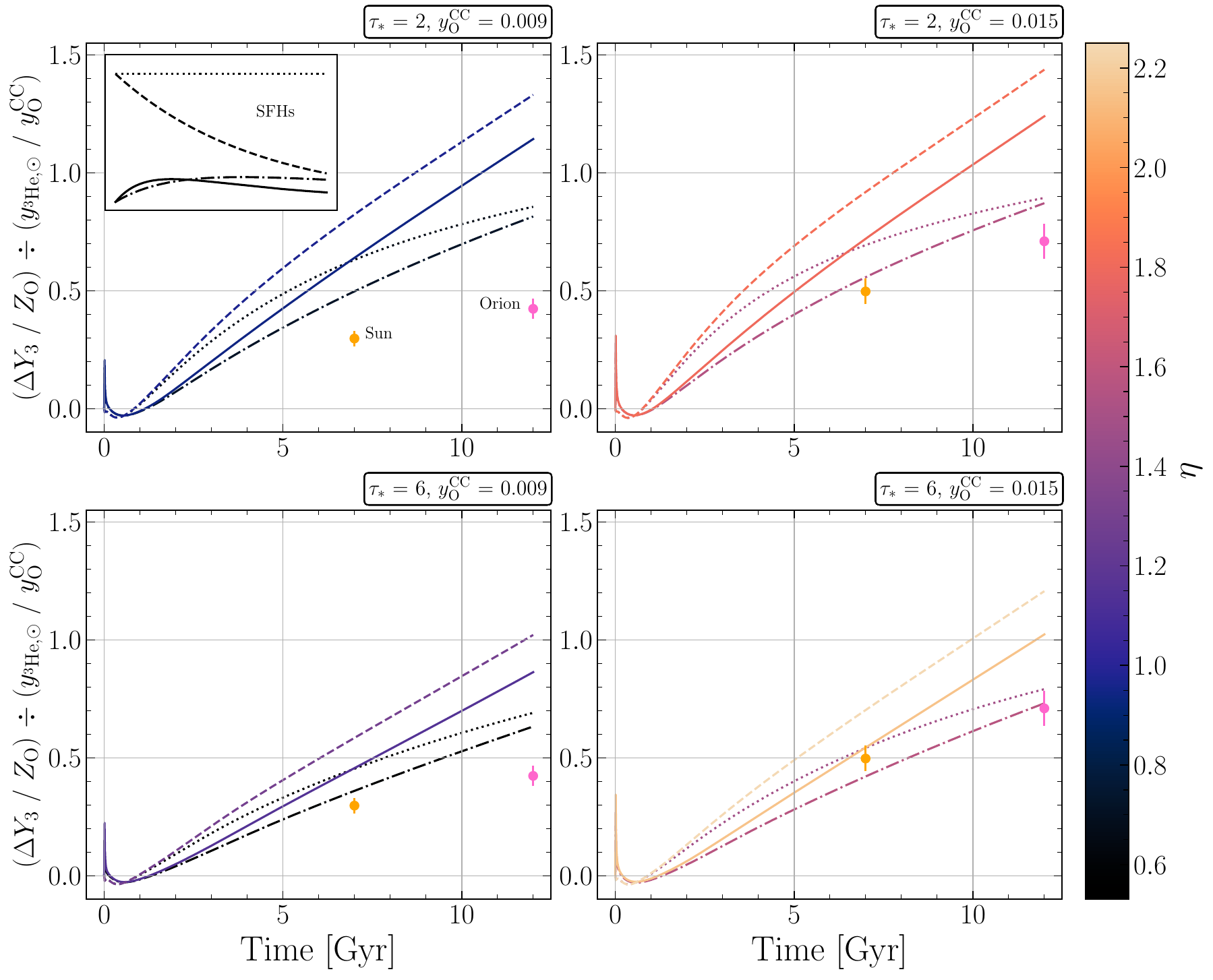}}
    \caption{Following the results from Figure \ref{fig: oz_vary}, we now constrain $\zo$ to achieve $\zosun$ at $t = 12\ \gyr$ by changing $\eta$ (colorbar). We also vary the oxygen yields (left-to-right) in addition to $\tau_*$ (top-to-bottom) and the SFH (linetypes). The $(\tau_1, \tau_2) = (4, 20)$ (dot-dashed) and $(\tau_1, \tau_2) = (0, \infty)$ (dotted) models reproduce the Solar (orange) and Orion (pink) abundances when $\yocc = 0.015$ only. Note that observational points move up on the right because they are normalized to the chosen oxygen yield of the model.}
    \label{fig: oz}
\end{figure*}

Figure \ref{fig: oz} examines the same four SFH models with $\tstar = 2\ \rm{Gyr}$ and $\tstar = 6\ \rm{Gyr}$, and it also considers models with a higher oxygen yield $\yocc = 0.015$ (right column). For each model, we choose a value of $\eta$ so that $\zo \approx \zosun$ at $t = 12\ \rm{Gyr}$. Curves in the upper left panel are nearly identical to those in the lower right panel of Figure \ref{fig: oz_vary}, but they differ slightly because we now adjust the value of $\eta$ from model to model. With $\tstar = 6\ \rm{Gyr}$ (lower left), the $(\tau_1, \tau_2) = (4, 20)$ model nearly matches the protosolar $\dely_3/\zo$ at $t = 7\ \rm{Gyr}$, though it still overpredicts the present-day $\dely_3/\zo$ measured in Orion. Models with $\yocc = 0.015$ require higher $\eta$ to reach $\zosun$ at late times. These larger outflows also eject $\het$, and the accreted gas that replaces the outflowing gas has the lower, primordial abundance. The effects on $\zo$ and $\dely_3$ nearly compensate, so model curves for $\dely_3/\zo$ are very similar between the left and right columns. However, the location of the data points shifts upward because the observed $\left(\dely_3/\zo\right)$ is normalized by a lower $\left(y_{\het}/\yocc\right)$. With the higher oxygen yield, the $(\tau_1, \tau_2) = (4, 20)$ and $(\tau_1, \tau_2) = (0, \infty)$ models are an acceptable match to the data for $\tstar = 6\ \rm{Gyr}$ and moderately overpredict the data for $\tstar = 2\ \rm{Gyr}$. 

\subsection{Two Infall Models}
\label{sec: two_infall}

\begin{figure*}
    \centering
\centerline{\includegraphics[width=0.7\paperwidth]{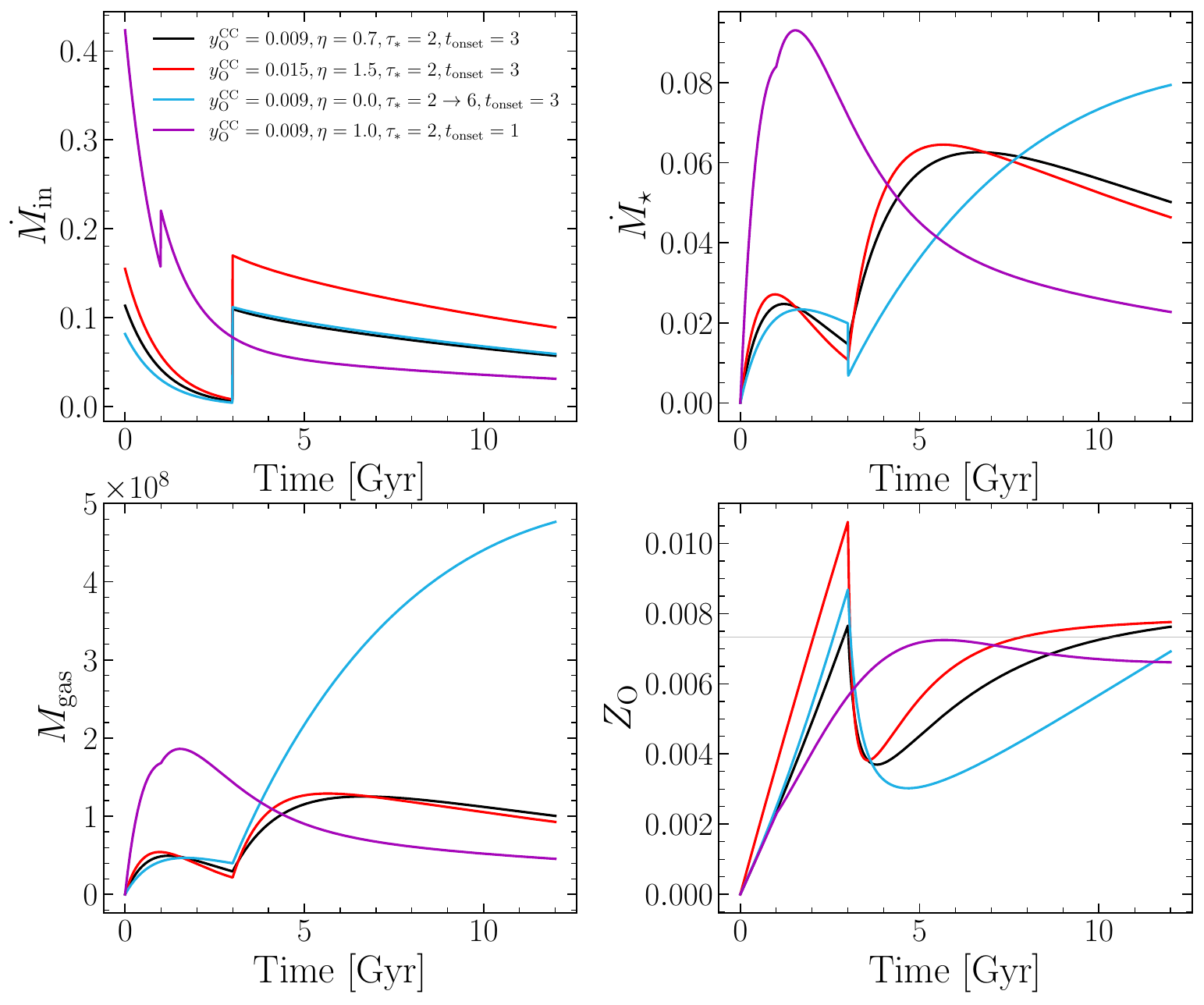}}
    \caption{Literature chemical evolution models of $\het$ use the two-infall model. Here, we run a suite of two-infall models changing the oxygen yields, outflows, onset time of the second infall, and the star-formation efficiency timescale shown as different colors. We plot the inflow rate, star-formation rate, gas mass, and oxygen gas abundance in different panels for clarity. The gray horizontal line in the bottom right panel represents $\zosun$.}
    \label{fig: twoinf}
\end{figure*}

The chemical evolution study of \citet{Lagarde2012} adopted the two-infall model \citep{Chiappini1997, Chiappini2001, Chiappini2002}, in which the high-$\alpha$ (``thick disk'') and the low-$\alpha$ (``thin-disk'') stellar populations are associated with distinct episodes of gas accretion separated in time. We have therefore run two-infall models in \vice, with parameters adapted from those of \citet{Dubay25}. Figure \ref{fig: twoinf} shows the gas infall history, star formation history, gas mass evolution, and $\zo$ evolution for the four two-infall models that we consider. Three of our models start the second infall episode at $t_{\rm onset} = 3\ \rm{Gyr}$ as advocated by \citet{Spitoni2019, Spitoni2020}, and the rapid gas dilution leads to a sudden drop in $\zo$, followed by slow recovery. We consider one model with $t_{\rm onset} = 1\ \rm{Gyr}$, which exhibits smoother evolution. In one model, we change the SFE timescale from $\tstar = 2\ \rm{Gyr}$ to $\tstar = 6\ \rm{Gyr}$ at $t_{\rm onset}$, and one model adopts the higher oxygen yield normalization $\yocc$. In all cases, we choose $\eta$ so that the model evolves to near solar metallicity at $t = 12\ \gyr$.

\begin{figure}
\centering
\centerline{\includegraphics[width=0.9\columnwidth]{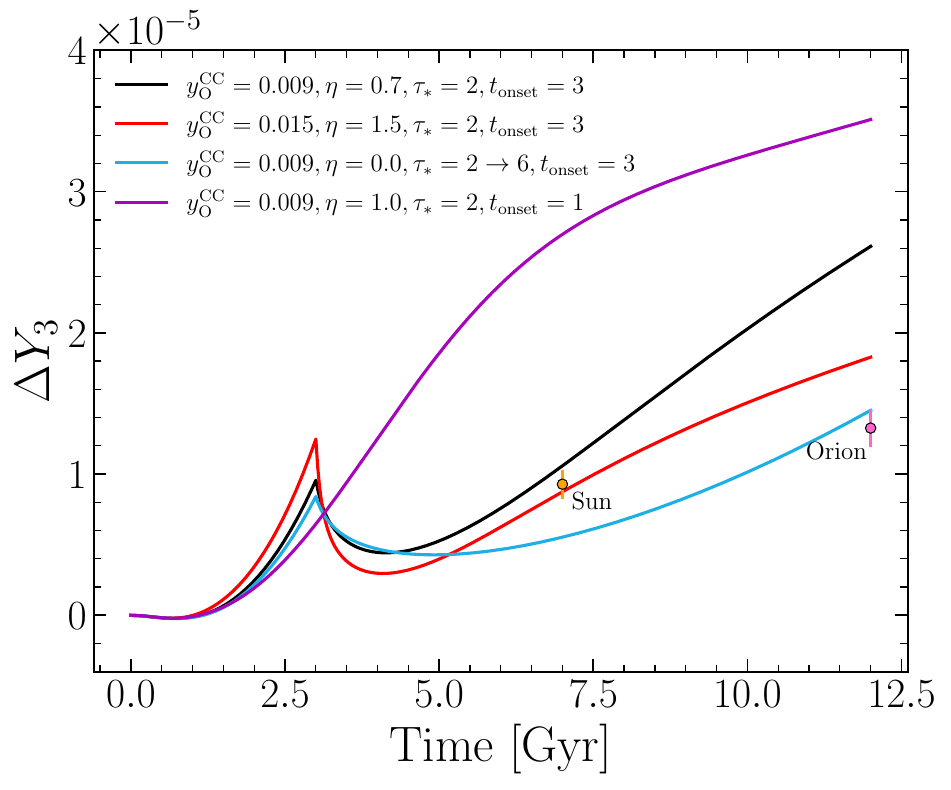}}
    \caption{$\het$ as a function of time for our two-infall models. By construction, the gas abundance is diluted and then built back up, more closely reproducing the Solar and Orion abundances of $\het$.}
    \label{fig: twoinf_3he}
\end{figure}

Figure \ref{fig: twoinf_3he} shows the $\dely_3$ evolution of these four models, with the protosolar and Orion measurements shown for comparison. With $t_{\rm onset} = 3\ \gyr$, dilution at the start of second infall leads to a sharp drop in $Y_3$, similar to the behavior of $\zo$. For $t_{\rm onset} = 1\ \gyr$, the evolution of $Y_3$ is smooth, but the model overpredicts the data. Two of these models come reasonably close to the data points: the model with high $\yocc$ normalization has high outflows ($\eta = 1.5$) that reduce $Y_3$, and the model with low SFE in the second infall phase evolves slowly and remains at low $Y_3$ even at late times. We have not found a model that has smooth $\dely_3$ evolution and comes reasonably close to the protosolar abundance, as one of the \citet{Lagarde2012} models does (their Figure 7), though even that model over predicts the protosolar $Y_3$ and would overpredict the Orion measurement more severely. Many details of the \citet{Lagarde2012} models are not specified in their paper, so we have not been able to track down the source of this difference.

\subsection{Summary}
\label{sec: one_zone_sum}

We can draw a number of lessons from these one-zone models. First, the long timescale of AGB $\het$ enrichment typically leads to steadily rising $Y_3$, in contrast to elements such as O, Fe, N, and $\hef$, which typically approach equilibrium at late times \citep[see, e.g.,][]{WAF, Johnson23, Weller2024}. Second, even with the lowest of the L11 AGB yields, for stellar models with thermohaline + rotational mixing, our models tend to overpredict the protosolar $Y_3$ and the present-day $Y_3$ measured in Orion. Third, models predict lower $Y_3$ in better agreement with data if the SFE timescale is long, leading to slower evolution, or the oxygen yield normalization is high, which implies (when normalizing to match the solar oxygen abundance) a high outflow efficiency that also removes synthesized $\het$ from the ISM. Finally, models can achieve better match to observed $Y_3$ values if the evolution has a slow or delayed start, as in our rise-fall model with $\tau_1 = 4\ \gyr$ or a two-infall scenario in which dilution at second infall resets $Y_3$ towards the primordial abundance.  

%% file: 4MultiZone.tex
\section{Multi-zone Galactic Chemical Evolution}
\label{sec: multizone}
With the results of our one-zone models as guidance, we now construct multi-zone GCE models designed to represent the evolution of the MW disk. We again use \vice, with parameter choices similar to those of \citet{Johnson21}, but adjusted to the lower oxygen yield scale used here, and with some parameter choices designed to give slower $\het$ evolution. \vice\ models the disk as a series of annuli \citep{Matteucci1989}, with $\tstar$, $\eta$, and SFH dependent on Galactocentric radius. Stars migrate from their birth radius, so at late times the stars present at a given radius are a mix of stars born at different radii \citep{Schonrich2009, Minchev2013, Minchev2014}. \citet{Johnson21} show that their models reproduce many of the observed features of the stellar age and abundance distribution of the MW, though the [$\alpha$/Fe] distribution is less bimodal than observed \citep[see also][]{Dubay25, Johnson25, Weller2024}.

\begin{figure}
\centering
\centerline{\includegraphics[width=0.9\columnwidth]{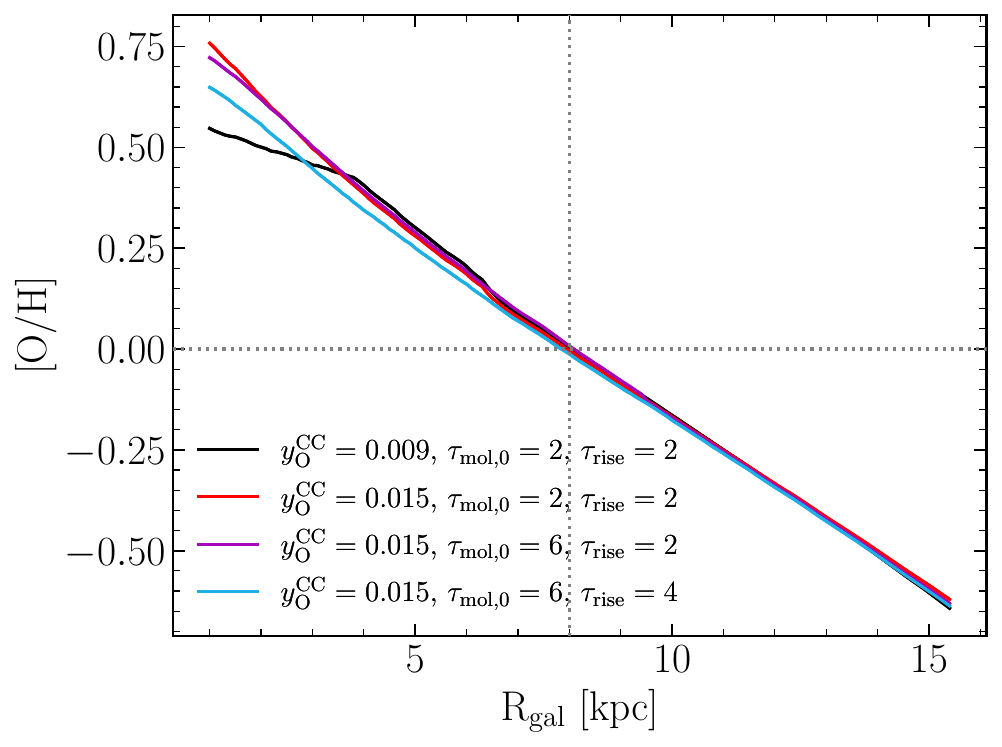}}
    \caption{Multizone model with changes to oxygen yields, and thus outflows, as well as star-formation efficiencies and rise timescales. The models are calibrated such that the oxygen gradient is nearly the same.}
    \label{fig: oh_grad}
\end{figure}

We calibrate our models to reproduce the observed MW [O/H] gradient at the present-day by setting the outflow efficiency to
\begin{equation}
    \label{eta}
    \eta (R_{\mathrm{gal}}) = \frac{y_{\alpha}^{\mathrm{CC}}}{Z_{\alpha, \odot}} 10^{(0.08 \mathrm{kpc}^{-1})(R_{\mathrm{gal}} - 4 \mathrm{kpc}) - 0.3} + r - 1.
\end{equation}
Our two choices of $\yocc$ from Section \ref{sec: onezone}, 0.009 and 0.015, correspond to $\yocc/\zosun = 1.23$ and $2.05$, respectively, and we set $r = 0.4$ in both cases. While the arguments that lead to Equation \ref{eta} are approximate, Figure \ref{fig: oh_grad} shows that our numerical models with two different choices of $\yocc$ and two different choices of SFE indeed predict nearly identical [O/H] gradients with a slope of about 0.08 dex/kpc and [O/H] = 0 at 8 kpc, in good agreement with observational estimates \citep{Frinchaboy2013, Hayden2014, Weinberg2019}. All of our models adopt the thermohaline + rotational mixing $\het$ yields of L11 and the LC18 massive star yields, as in Section \ref{sec: onezone}. 

Instead of a constant $\tstar$, the \citet{Johnson21} models adopt a star formation law (their Equation 14) in which the SFE increases with gas surface density \citep{Kennicutt1998, Krumholz2018} but saturates to a constant value, $\dot\Sigma_*/\Sigma_{\rm gas} = \tau_{\rm mol}^{-1}$ for $\Sigma_{\rm gas} > 2 \times 10^{7} \M\ \rm{kpc}^{-2}$, in agreement with observations implying roughly constant SFE in molecular gas \citep{Leroy2008, Sun22}. Following \citet{Johnson21}, we set
\begin{equation}
    \label{tmol}
    \tau_{\rm mol} = \tau_{\rm mol, 0} \left(t / t_0\right)^{1/2}
\end{equation}
with $\tau_{\rm mol, 0} = 2\ \gyr$ and a time dependence motivated by the observations of \citet{Tacconi2018}. We also consider low SFE models with $\tau_{\rm mol, 0} = 6\ \gyr$. 

\citet{Johnson21} adopt a rise-fall SFH of 
\begin{equation}
    \label{sfr}
    \dot M_* = (1 - e^{-t / \tau_1})e^{-t / \tau_2}.
\end{equation}
with rise time $\tau_1 = 2\ \gyr$ and a decline timescale $\tau_2$ chosen to reproduce observed age profiles of stellar disks from \citet{Sanchez2020}. We also consider a model with a slower rise of star formation, $\tau_1 = 4\ \gyr$. Because the rise-fall SFH allows a slow onset of star formation, we evolve these models for 13.2 Gyr rather than the shorter, 12 Gyr disk lifetime used in Section \ref{sec: onezone}.

\begin{figure}
\centering
\centerline{\includegraphics[width=0.9\columnwidth]{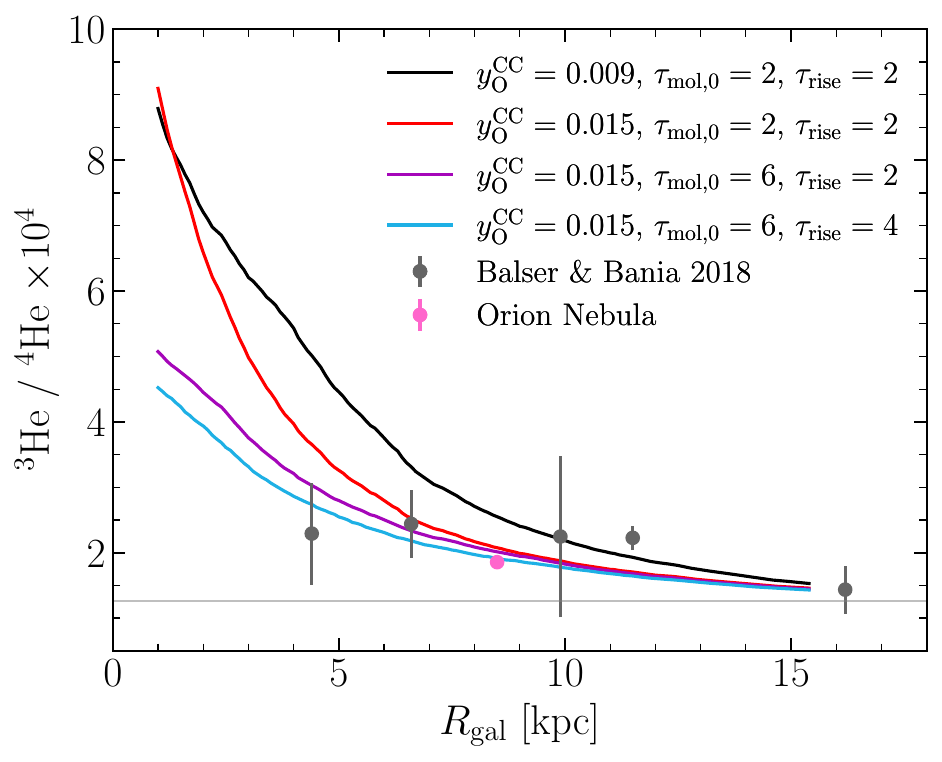}}
    \caption{Present-day, gas-phase radial gradient of $\het$ to $\hef$ for our multizone models shown as curves. We use the $\hef$ yields advocated by \citet{Weller2024}. The models with $\yocc = 0.015$ adequately predict the Orion abundance (pink) at $R = 8.5\ \rm{kpc}$ as well as the observations of H\,\textsc{ii} regions in the MW from \citet{Balser2018} (gray).}
    \label{fig: 3he_grad}
\end{figure}

Figure \ref{fig: 3he_grad} plots the present-day gas-phase gradient of $\het/\hef$ for our four multi-zone models. For $\hef$, with our fiducial $\yocc = 0.009$, we adopt the yields advocated by \citet{Weller2024}, which combines the NKT13 massive star yields for $8 \M \leq m \leq 40 \M$ and the LC18 yields at $m > 40 \M$ to capture wind and explosive contributions of more massive stars. When running models with $\yocc = 0.015$, we adopt the higher massive star $\hef$ yields that LC18 find for rapidly rotating stars, with initial rotational velocities of $300 \kms$. In both cases, we use the AGB yields of \citet{V13}. The $\hef$ yields have a small effect on the $\het/\hef$ predictions, since the evolved $\hef$ is only $\sim 10\%$ above the primordial value. For observational data, we take the Orion measurement of \citeauthor{Cooke22}~(\citeyear{Cooke22}, \citeyear{Cooke2026}) at $R = 8.5\ \rm{kpc}$, and we convert \citet{Balser2018} measurements of $\het/{\rm H}$ to $\het/\hef$. Specifically, we use their Equation 2 to calculate $y_4$, the $\hef$ abundance by number, and divide their reported $\het/\rm{H}$ by this value. The precision of the \citeauthor{Cooke22}~(\citeyear{Cooke22}, \citeyear{Cooke2026}) measurement is substantially higher than that of the \citet{Balser2018} measurements, and it is likely more robust because it measures the same $\lambda \approx 10830 \textup{~\AA}$ absorption line for $\het$ and $\hef$. 

The model with the parameter choices we consider best motivated by other considerations, $\yocc = 0.009$, $\tau_{\rm mol, 0} = 2\ \gyr$, $\tau_1 = 2\ \gyr$, clearly overpredicts the Orion data, and it also overpredicts the \citet{Balser2018} points at smaller \rgal\ (black curve). As expected from the one-zone models of Section \ref{sec: onezone}, raising the oxygen yield to $\yocc = 0.015$ (red curve) leads to better agreement, since higher outflows (Equation \ref{eta}) eject some of the $\het$ produced by AGB stars from the ISM. This model still overpredicts the Orion measurement, and better agreement is obtained by increasing the SFE timescale to $\tau_{\rm mol, 0} = 6\ \gyr$ (purple curve) and by further increasing the SFH rise time to $\tau_1 = 4\ \gyr$ (blue curve). At large $R$, the predicted abundance ratios approach the primordial value but remain above. The best discrimination between models would come from precise measurements at smaller radii, where enrichment is higher. 

\begin{figure}
\centering
\centerline{\includegraphics[width=0.9\columnwidth]{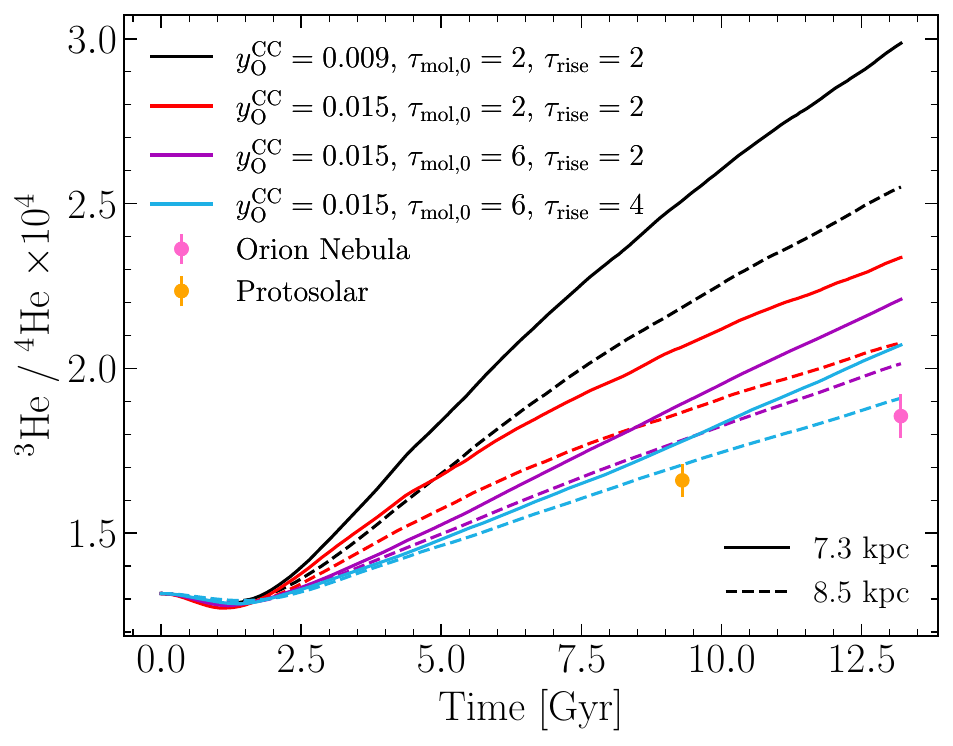}}
    \caption{$\het/\hef$ ratio as a function of time in our models compared to measurements of the protosolar abundance \citep[orange,][]{Mahaffy1998} and the Orion nebula abundance (pink, \citeauthor{Cooke22}~\citeyear{Cooke22}, \citeyear{Cooke2026}). Dashed curves show results at 8.5 kpc, for comparison to Orion, and solid curves show results at 7.3 kpc, an estimate of the solar birth radius.}
    \label{fig: 3he_time}
\end{figure}

Figure \ref{fig: 3he_time} plots the time evolution of these models at $R = 8.5\ \rm{kpc}$, corresponding to Orion, and $R = 7.3\ \rm{kpc}$, which we take as the solar birth radius based on \citet{Minchev18}. We note, however, that the solar birth radius is uncertain and could plausibly be in the range $4 - 9\ \rm{kpc}$ \citep{Nieva12, Frankel2018, Tsujimoto20, Baba23}. Since all our models overpredict the solar $\het/\hef$, ratio, at least slightly, a smaller birth radius would exacerbate the discrepancy and a larger birth radius would reduce it. Taking the observational error bars and solar birth radius estimate at face value, only the model with all parameters chosen to minimize $\het/\hef$ comes close to the data. The model with $\yocc = 0.009$, $\tau_{\rm mol, 0} = 2\ \gyr$, $\tau_1 = 2\ \gyr$, which would be our preferred model based on other considerations, overpredicts the observed $\het/\hef$ by a factor of $\sim 1.4$. 

In our \vice\ models, outflow is implemented as simple ejection from the ISM. However, from the point of view of chemical evolution, a ``fountain'' flow with re-accretion of ejected gas is nearly equivalent to pure outflow \textit{if} gas that is ejected from the disk is mixed with a much larger amount of primordial gas before returning, diluted, to the ISM \citep[see Section 5.2 of][]{WAF}. Conversely, a fountain flow that returns undiluted is nearly equivalent to no outflow. The effect of radial gas flows resembles that of outflows in many respects \citep{JohnsonRadialFlows}, and it might be possible to construct successful $\het$ evolutionary models with radial flows instead of outflows. However, models with strong radial flows would still need high $\yocc$ to reproduce the observed [O/H] gradient. 

%% file: 5Conclusions.tex
\section{Conclusions}
\label{sec: conclusions}

We have constructed models for the evolution of $\het$ in the MW disk, comparing our results to the estimated protosolar abundance \citep{Mahaffy1998}, the measurement of $\het/\hef$ in the Orion nebula by \citeauthor{Cooke22}~(\citeyear{Cooke22}, \citeyear{Cooke2026}), and lower precision measurements of $\het/\rm{H}$ in the ISM by \citet{Balser2018}. With the massive stars models of LC18, NKT13, or S16, we find that the IMF-averaged net yields of $\het$ are negligible compared to the BBN primordial abundance and to the AGB yields predicted by L11. The L11 yields are metallicity dependent, and at $Z = Z_{\odot}$ the predicted IMF-averaged, net yield for ``standard'' stellar models is $\yheagb = 8.47 \times 10^{-5}$ (solar masses of $\het$ produced per solar mass of stars formed). This yield is larger than both the protosolar $Y_{3, \odot} = 3.38 \times 10^{-5}$ and the primordial $Y_{\rm 3, P} \approx 2.45 \times 10^{-5}$ predicted by BBN \citep[e.g.,][]{Yeh2021, Pitrou21}. However, thermohaline and rotational mixing can reduce the net $\het$ production by bringing $\het$ from the stellar envelope to layers where it can be destroyed by fusion. With both mixing processes included, the L11 yields drop by a factor of two or more, with $\yheagb = 3.82 \times 10^{-5}$ at $Z = Z_{\odot}$. Our IMF-averaged yield comparisons are summarized in Figure \ref{fig: imf_yields}. 

With or without extra mixing, the L11 yields are dominated by stars with $m = 1 -2 \M$, which have main sequence lifetimes $\gtrsim 1\ \gyr$. As a result, the delay time distribution for $\het$ enrichment is extended, with half of the production coming at $t \gtrsim 2\ \gyr$ (Figure \ref{fig: agbdelay}). This behavior contrasts with that of other AGB elements we have previously examined such as Sr \citep{Johnson20}, N \citep{Johnson23}, or $\hef$ \citep{Weller2024}, where production is dominated by higher mass stars with shorter lifetimes. The slow production of $\het$ makes its predicted evolution qualitatively different from that of these other AGB elements. The difference is rooted in  basic nucleosynthesis: low mass stars produce more $\het$ because of the dominance of pp hydrogen burning over the CNO bi-cycle and longer main sequence lifetimes \citep{Lagarde2011}.

In \citet{Weller2024}, we showed that the evolution of $\dely = Y - Y_{\rm P}$ is similar to that of promptly produced elements such as O or Mg, with small differences because of delayed AGB prediction and metallicity-dependent yield. For $\dely_3 = Y_3 - Y_{3, \rm P}$, on the other hand, long AGB time delays lead to much slower evolution, so the ratio $\dely_3/\zo$ tends to increase steadily in time (Figure \ref{fig: oz_vary}). Even with the lowest (thermohaline + rotational mixing) yields from L11, our one-zone GCE models tend to overpredict the observed $Y_3$ values, so we do not consider the other L11 yields. The predicted $Y_3$ is lower for higher outflow efficiency $\eta$, which ejects more synthesized $\het$ from the ISM, for longer SFE timescale $\tstar$, which leads to slower evolution, or for a slower onset of star formation, as in our rise-fall SFH models (Figure \ref{fig: oz_vary}). 

We adopt $\yocc = 0.009$ as our fiducial oxygen yield, based on the arguments of \citet{Weinberg24} and \citet{Weller2024}. When we choose $\eta$ so that our one-zone GCE models reach $\zo \approx \zosun$ at late times, they overpredict the protosolar and Orion $Y_3$ values. If we increase $\yocc$ to 0.015, then the $\eta$ values are higher, and models that also adopt low SFE ($\tstar = 6\ \gyr$) and an SFH that is nearly constant (instead of declining) at late times come close to the observed $Y_3$ values (Figure \ref{fig: oz}). 

We also examined one-zone two-infall models, and we find that the ``reset'' of ISM abundances at the start of the second infall leads to lower $Y_3$ values in better agreement with observations (Figure \ref{fig: twoinf_3he}). In particular, we find a reasonably successful model with $\yocc = 0.009$ when we combine a second infall with a transition to lower SFE. However, these models predict a substantial rise in $\zo$ over the past 8 Gyr, in tension with evidence for a stellar abundance gradient that is steady in time \citep[see][]{Dubay25, Johnson25}.

Turning to multi-zone models like those of \citet{Johnson21}, we again find that models with $\yocc = 0.009$ lead to overproduction of $\het$ with the L11 yields. A higher $\yocc = 0.015$ leads to a higher normalization of $\eta(R)$ and better agreement with observed $\het/\hef$. Our most successful multi-zone model has $\yocc = 0.015$, low SFE with $\tau_{\rm mol, 0} = 6\ \gyr$, and a slow onset of star formation (Figures \ref{fig: 3he_grad} and \ref{fig: 3he_time}). While our models differ from those of \citet{Cooke22} in many details, they agree on the qualitative conclusion: reproducing observed $\het/\hef$ ratios with the L11 stellar yields requires substantial outflows from the ISM. Note that in our models, the outflow efficiency is linked to $\yocc$ by the requirement of matching the observed [O/H] gradient, and higher outflow implies higher inflow because we fix the SFH and thus $\Sigma_{\rm gas} (R, t) = \dot{\Sigma_*} (R, t)/\tstar(R,t)$. 

Recently, \citet{Cooke2026} have presented two new measurements of the ISM $\het/\hef$ ratio and multi-zone \vice\ models constrained by these measurements. In contrast to our approach here, \citet{Cooke2026} take the primordial $\het/\hef$ ratio and the scale of $\yocc$ as fitting parameters. Not surprisingly, these two parameters are significantly degenerate, and for $Y_{3, \rm P}$ equal to the standard BBN prediction, they infer $\yocc/\zosun \approx 2 - 3$, also adopting the L11 thermohaline + rotational mixing AGB yields. We have instead assumed either $\yocc = 0.009 \approx 1.2 \zosun$ or $\yocc = 0.015 \approx 2 \zosun$ and investigated a wider range of other parameters in our one-zone and multi-zone models. \citet{WeinbergDH}, \citeauthor{Cooke22}~(\citeyear{Cooke22}, \citeyear{Cooke2026}), \citet{JohnsonIsotopes}, and this paper have all emphasized the value of D and $\het$ as diagnostics for the overall scale of nucleosynthetic yields and the importance of outflows in MW chemical evolution. In all cases, these analyses favor higher yields and outflows, which are also favored by models that match the slow evolution of the MW metallicity gradient \citep{Dubay25, JohnsonRadialFlows, Johnson25}, but which are difficult to reconcile with empirical estimates of the mean CCSN iron yield \citep{Weinberg24}.

For D, the principal uncertainty is the uncertain value of the ISM D/H ratio \citep{Linsky2006}, which is measured along a relatively small number of sightlines and strongly affected by dust depletion. For $\het$, the principal uncertainty is the theoretical stellar yield. We find that even the lowest of the L11 yields requires straining other chemical evolution parameters to match the protosolar and Orion values of $Y_3$, and the ``standard'' models without extra mixing would be very difficult to reconcile with observed $\het$ abundances (as already concluded by \citet{Lagarde2012}). With the prospect of further $\het/\hef$ measurements using the methods of \citeauthor{Cooke22}~(\citeyear{Cooke22}, \citeyear{Cooke2026}), a new investigation of AGB yields would be warranted to see whether yields can be significantly lower than those of the L11 thermohaline + rotational models. From an empirical point of view, measurements that probe super-solar and sub-solar ISM metallicities would be valuable, as these could confirm or refute the trends predicted by chemical evolution models for given yield choices. The large primordial contribution to $\het$ combined with slow stellar production dominated by long-lived stars make it a unique probe of the MW's enrichment history.

%% file: 6Appendix.tex
\section{Analytic Solutions}
\label{appendix:extras}

Our analytic solutions (Figure \ref{fig: cz_vs_dlz}) are adapted from those of \citet{WAF}. Assuming an exponentially declining star formation rate, the equilibrium solution for oxygen is given by:
\begin{equation}
    \label{zoeq}
    Z_{\mathrm{O, eq}} = \frac{m_{\rm O}^{\rm CC}}{1 + \eta - r - \tstar/\tsfh}.
\end{equation}
The full time evolution is described by
\begin{equation}
    \label{oxy_analytic}
    \zo \left(t\right) = Z_{\mathrm{O, eq}} \left(1 - e^{-t/\bar{\tau}_{\rm [dep, SFH]}}\right),
\end{equation}
$\tau_{\rm dep} = \tstar / (1 + \eta - r)$ is the gas depletion timescale and $\bar{\tau}_{\rm [dep, SFH]} = \left(\tau_{\rm dep}^{-1} - \tsfh^{-1}\right)^{-1}$. The smaller of $\tau_{\rm dep}$ and $\tsfh$ controls the rate of evolution, unless they are close in magnitude. \\
For $\het$, the approach to equilibrium is
\begin{equation}
    \label{analytic}
    Y_{\het, \rm eq}^{\rm AGB} = \left(\frac{\bar{\tau}_{\rm [AGB, SFH]}}{\tau_{\rm AGB}}\right) e^{t_D/\tsfh} \frac{y_{\het}^{\rm AGB}}{1 + \eta - r - \tstar/\tsfh},
\end{equation}
where $t_D$ is the lifetime of an $8 \M$ star, indicating the start of AGB enrichment, and the enrichment timescale $\tau_{\rm AGB}$ is discussed in Section \ref{agb_yields}. Here $\bar{\tau}_{\rm [AGB, SFH]} = \left(\tau_{\rm AGB}^{-1} - \tsfh^{-1}\right)^{-1}$. For $\tau_{\rm AGB} = 3.71\ \gyr$ (L11 thermohaline + rotational yields) and $\tsfh = 6\ \gyr$, the rate of evolution for $\het$ is significantly longer than $\tau_{\rm AGB}$ or $\tsfh$, because time-delayed AGB enrichment enters a declining gas supply. \\
The full time evolution of $\het$ is
\begin{equation}
\label{analytic_time}
\begin{aligned}
Y_{\het}^{\rm AGB}(t)
&= Y_{\het, \rm eq}^{\rm AGB}
\Bigg[
1 - e^{-\Delta t / \bar{\tau}_{\rm [dep, SFH]}} \\
&\quad
- \frac{\bar{\tau}_{\rm [dep, AGB]}}{\bar{\tau}_{\rm [dep, SFH]}}
\left(
e^{-\Delta t / \bar{\tau}_{\rm [AGB, SFH]}}
- e^{-\Delta t / \bar{\tau}_{\rm [dep, SFH]}}
\right)
\Bigg].
\end{aligned}
\end{equation}
The \citet{WAF} solutions are derived for a metallicity-independent yield. When applying Equations \ref{analytic} and \ref{analytic_time}, we either set $y_{\het}^{\rm AGB}(t)$ equal to the value for metallicity $Z/Z_{\odot} = \zo(t)/\zosun$ or (in our ``Delayed Z'' model) to the value for $Z/Z_{\odot} = \zo(t - \tau_{\rm dep})/\zosun$. Analytic solutions for yields that depend linearly on metallicity are derived by \citet{Sanders25}.